\documentclass[a4paper,11pt]{article}
\pdfoutput=1 % if your are submitting a pdflatex (i.e. if you have
\usepackage{jcappub} % for details on the use of the package, please
\usepackage[T1]{fontenc} % if needed

\usepackage{subcaption}

\usepackage{adjustbox}
\usepackage{booktabs} 
\title{\boldmath Reheating and Inflationary dynamics driven by an inverse tangent potential}

\author[a]{Mayur Abhisheki,}
\author[a,1]{Prasanta Kumar Das,\note{Corresponding author.}}

\affiliation[a]{Department of Physics, Birla Institute of Technology and Science, \\
K. K. Birla Goa Campus, NH-17B, Zuarinagar, Sancoale, Goa 403726, India}

\emailAdd{p20240040@goa.bits-pilani.ac.in}
\emailAdd{pdas@goa.bits-pilani.ac.in}

\abstract{In this work, we study the early universe inflation and the post-inflation reheating era employing an inverse tangent potential of the form $V=V_0 \cdot[tan^{-1}(\frac{\kappa \phi}{m_p})]^2$, where $\kappa$ is a free parameter of the potential and $m_p$ is the reduced Planck mass. We derive the slow roll parameters, the number of e-folds(N), the scalar spectral index $n_s$, the tensor-to-scalar ratio $r$, and the tensor spectral index $n_T$ for the inverse tangent potential. We examine the inflationary observables using the data of the \emph{Planck}-2018 and  recent ACT collaboration and obtain constraints on the potential parameter $\kappa$.  We also employ a reheating analysis by invoking the conservation of entropy between today and the time when reheating starts. We obtain bounds on the reheating temperature $T_{re}$ and the number of e-folds of the reheating $N_{re}$ using the spectral-index $n_s$ constraints from \emph{Planck} 2018 and the ACT results. We show that this inverse-tangent potential can act as an alternative to the standard inflationary potentials like Starobinsky which are excluded at $2\sigma$ level by the recent sixth data release (DR6) of the Atacama Cosmology Telescope (ACT) collaboration.}

\begin{document}
\maketitle
\flushbottom

%%%%%%%%%%%%%
\section{Introduction}

In 1970s Big bang cosmology seemed a very successful theory but by the end of the decade two problems emerged, and they were the Horizon and the Flatness problems. A. Guth proposed the idea of inflation now called the 'Old Inflation' \cite{Guth} in order to solve the monopole problem, which also seemed to solve the horizon and flatness problems. However, this model was flawed because the inflaton is initially trapped in a false vacuum. As the inflaton leaks through the potential barrier and forms bubbles of true vacuum, the energy released in the transition ends up concentrated within the bubble walls. If the bubbles are able to merge, a homogeneous and isotropic universe emerges. However, the bubbles never collide, since the background false-vacuum space in which they formed, never stops inflating. However, this exit problem can be solved by ensuring that the potential is sufficiently flat post-nucleation, thus replacing the need for false vacuum and this is given the name 'New Inflation' \cite{Linde},\cite{Albrecht}.   \\
Although not aware at the time, Starobinsky had proposed a model of inflation \cite{Starobinsky} through an $R^2$ addition to the Einstein-Hilbert action. Since then a multitude of inflationary potentials have been proposed, motivated by both phenomenology and high-energy theory like chaotic inflation \cite{LindeC}, Higgs Inflation \cite{Higgs}, Natural inflation \cite{Natural} etc. It was soon realized that the quantum fluctuations during inflation can account for the density perturbations necessary for growth of structure in the universe \cite{Hawking,Starobinsky2,Guth2,Bardeen}. Measurements of the temperature fluctuations of the Cosmic Microwave Background (CMB) by the recent \emph{Planck} collaboration \cite{Planck} have provided strong observational support for the inflationary scenario. 
Also in the recent sixth data release (DR6) of the Atacama Cosmology Telescope (ACT) collaboration \cite{ACT,ACT2} reported new constraints on spectral index $n_s$ and other cosmological parameters which indicate exclusion of potentials like  the Starobinsky and Higgs inflationary models at $2\sigma$ level.
The straightforward way to study inflation is to consider that inflation is driven by a scalar field called Inflaton, which under the influence of a particular potential along with the slow-roll approximation (where the kinetic terms are neglected with respect to the potential term) is used in order to examine the inflationary scenario.\\
Once inflation ends, the inflaton field begins to decay into standard model particles, initiating the production of ordinary matter. This transition period, bridging the inflationary phase and the onset of radiation and matter domination, is known as reheating. The dynamics of reheating can vary significantly depending on the model. In some scenarios, the inflaton undergoes perturbative decay \cite{Abbott,Dolgov,Albrecht2}, while others involve non-perturbative mechanisms such as broad parametric resonance (preheating) \cite{Kofman, Traschen, Kofman2}, tachyonic instabilities \cite{Greene1997, Shuhmaher2006, Dufaux2006, Abolhasani2010, Felder2001a, Felder2001b}, or instant preheating \cite{Felder1999}.

In this work, we have investigated the non-polynomial inverse tangent potential - which was less explored, but still yielded interesting phenomenology. This type of potential was used in cosmological parameter estimation and investigation of spectra in the slow-roll expansion \cite{Leach}, and also in the study of formation of Primordial Black Holes (PBHs) in single-field models \cite{Drees}. The inverse tangent model was also used to study the  thermodynamics of a black-hole solution in 2D dialton gravity \cite{Belhaj}, in the study of brane inflation \cite{Belhaj2} and also in the study of the non-minimal coupling with gravity in presence of an inflaton field \cite{Naciri}. Such an inverse tangent potential was obtained in a cosmological model, through brane inflation scenario \cite{Neves}.

In the context of this paper, first we employ a slow-roll analysis to this potential and obtain different observable quantities like spectral index $n_s$, tensor-to-scalar ratio $r$ and analyze their behavior with respect to the potential free parameter $\kappa$.
We determine the running of the spectral index ($n_s$) and estimated the non-Gaussianity parameter $f_{NL}$. Using the \emph{Planck} 2018 and ACT collaboration results of $\alpha_s$, $n_s$, $r$ and $f_{NL}$, we obtain constraints on the potential parameter $\kappa$.  We then move to reheating. During the reheating era, the energy density is assumed to vary with the scale factor($a(t)$) as $\rho\propto a^{-3(1+\omega)}$, where $\omega$ is the equation of state parameter. One requires $\omega>-1/3$ to end inflation and the numerical studies of the thermalization have suggested a range of variation of $0<\omega<0.25$ \cite{Dai, Munoz, Cook, Ueno}. It has been demonstrated that such an analysis can be utilized to eliminate degeneracies resulting from various inflationary models \cite{Swagat}. For the purposes of this work we will thus approximate $\omega$ as a constant in all our calculations. We determine the number of e-fold $N_{re}$ and $T_{re}$ which are the duration of reheating and the thermalization temperature, respectively. For the potential slow-roll as well as reheating analyses, we compare our model with the best-fitting {\it Starobinsky model}. 

The paper is organized as follows: In Section (\ref{Our Model}), we define model and outline the necessary theoretical framework. Section (\ref{Slow-Roll Analysis}) presents our calculation and plots from the slow-roll analysis. Section (\ref{Reheating Analysis}) discusses reheating and its constraints. Finally, in Section (\ref{Conclusion}), we summarize our findings and comment on future directions.

%%%%%%%%%%%%%%%%%%%%%%%%%%%%%%%%%%%%%%%%%%%%%%%%%%%%%%%%
\section{Our Model} \label{Our Model}

During inflation, the dominant contribution to the energy momentum tensor comes from the inflaton scalar field($\phi$) described by the Lagrangian \begin{equation}
\mathcal{L} = \frac{1}{2}\partial_\mu \phi \partial^\mu \phi - V(\phi)
\end{equation}
 The equations that govern  the dynamics of space-time expansion of the isotropic and homogeneous universe, are the Friedmann equation, the Raychaudhury equation, and the inflaton equation
\begin{equation}
  H^2 = \left(\frac{\dot{a}}{a}\right)^2 =\frac{1}{3m_p^2}\rho_{\phi}
    \label{fr}
\end{equation}
\begin{equation}
    \frac{\ddot{a}}{a} = -\frac{1}{6 m_p^2}(\rho_{\phi} + 3 p_\phi)
    \label{ry}
\end{equation}

\begin{equation}
    \ddot{\phi}+ 3H\dot{\phi} + V^{'}(\phi)=0
    \label{eom}
\end{equation}
 where $\rho_\phi=\frac{1}{2}\dot\phi^2+V(\phi)$, $p_\phi=\frac{1}{2}\dot\phi^2-V(\phi)$ and $m_p^2=\frac{1}{8\pi G}$.
 %=2.43\times10^{18}$ GeV the reduced Planck mass-squared.
Here an over-dot represents the derivative with respect to time ($t$) and the prime denotes derivative with respect to $\phi$. \\
\noindent Inflation, in principle, is driven by a scalar field $\phi$ whose potential is sufficiently flat, as measured by the potential slow roll parameters  
%In our case, we try to model it with a potential of the form
%\begin{equation}
%    V = V_0\cdot \left [ tan^{-1}(\frac{\kappa \phi}{m_p})\right ]^2
%    \label{pot}
%\end{equation}
%where $\kappa$ is a dimensionless free parameter and $V_0$ has dimensions of $[Energy]^4$.\\

\begin{equation}
    \epsilon=\frac{m_p^2}{2}\left ( \frac{V^{'}}{V} \right )^2,~ \eta=m_p^2 \left ( \frac{V^{"}}{V}\right)
    \label{sr}
\end{equation}
In the slow-roll approximation, we have $\epsilon <<1,~|\eta|<<1$. When $\epsilon, ~|\eta|<<1$, we can neglect the term $\ddot{\phi}$ in the equation of motion of the inflaton field $\phi$ can be ignored, and rewrite it as  
\begin{align}
     &3H\dot{\phi}\simeq -V^{'}(\phi)\\
     &3H^2\frac{d\phi}{dN}=-V^{'}(\phi)
     \label{H}
\end{align}
where we make use of the fact that $N=\int Hdt$ is the number of e-foldings. From Equation \ref{H}, we can write 
\begin{equation}
    N_k=-\int^{\phi_e}_{\phi_k} d\phi \frac{3H^2}{V^{'}} 
    \label{Ne}
\end{equation}
To find the number of e-folds, Equation \ref{Ne} can be integrated from the point when the pivot mode $k_*$ crossed the horizon ($\phi = \phi_*$) till the end of inflation ($\phi = \phi_{end}$).\footnote{In our analysis, the choice of this pivot scale is \emph{Planck}'s pivot scale of $k_*=0.05$ Mpc$^{-1}$}.

An important observable quantity is the 3D curvature correlation function, which can be understood from the scalar power spectrum $P_s$ \cite{Cline}
\begin{equation}
    P_s=\int d^3x e^{ikx} \langle \mathcal{R}(0)\mathcal{R}(x) \rangle =A_s\left(\frac{k}{k_*}\right                   )^{n_s-1}
\end{equation}
where $n_s$ is known as the spectral index and it quantifies the scale invariance of the power spectrum. The amplitude of the scalar power spectrum can then be constrained as 
\begin{equation}
    A_s=\frac{V(\phi)}{24\pi^2m_p^4\epsilon}=\frac{1}{12\pi^2m_p^6}\frac{V^3}{V'^2}
\end{equation}
The vector perturbations decay in the expanding space, and hence we can ignore them. Similar to the scalar perturbations, gravity waves get quantum fluctuations. However, these fluctuations are effective only at large scales, and the power spectrum associated with them is 
\begin{equation}
    P_t=A_t\left(\frac{k}{k_*}\right)^{n_t-1}
\end{equation}
Then the tenor-to-scalar ratio is given by \begin{equation}
    r=\frac{A_t}{As}
\end{equation}
The phase of inflationary expansion will cease to exist when the slow-roll parameter $\epsilon$ becomes equal to 1, characterized by the value of $\phi$ as $\phi_e$. In our domain of slow-roll, we also define certain observable quantities, the spectral index $n_s$, the tensor-to-scalar ratio $r$ and running of the spectral index $\frac{dn_s}{d lnk}$ as  

\begin{equation}
n_s \simeq -6\,\epsilon + 2\,\eta + 1,\quad
r \simeq 16\,\epsilon
\quad\text{and}\quad
\frac{d n_s}{d \ln k}
= 16\,\epsilon\,\eta - 24\,\epsilon^2 - 2\,\xi^2.
\label{si}
\end{equation}
\noindent where $\xi=m_p^4 \frac{V^{'}V^{'''}}{V^2}$. \\
\noindent Up to a very good approximation, the fluctuations from a standard inflation behave as Gaussian. But still, non-Gaussianity can arise if the primordial fluctuations are non-gaussian in nature. Although single-field inflation models demand that interactions of the inflaton field be weak \cite{nong},\cite{nong2} and the fluctuations created during inflation are expected to be gaussian as the amplitude of the bispectrum is suppressed by slow roll parameters. We try to verify this by calculating the non-linearity parameter $f_{NL}$ which is calculated as \cite{Koh},\cite{Wands}
\begin{equation}
    f_{NL}=\frac{5}{6}(\eta-3\epsilon)
\end{equation}
for single-field slow-roll inflation.

Once the slow-roll phase ends (around $\epsilon=1$), the inflaton has to lose all its energy and get converted to relevant degrees of freedom of the Standard Model. The inflaton here begins to oscillate about the minimum and decays to matter and radiation. This transition from inflationary phase which is cold to the end of inflation which is hot and radiation dominated is called reheating. This reheating process is essential for big-bang nucleosynthesis.

It has been seen that for a given interaction between the inflaton field and matter fields, we observe a relationship of the type

\begin{equation}
    p=\omega_{re}(t) \rho
\end{equation}
where p is pressure, $\rho$ is energy density, and $\omega_{re}(t)$ is an effective equation of the state parameter. In this paper we follow similar approach as in \cite{Dai, Munoz, Cook, Ueno} and use the fact that the number of e-folds of the reheating era can be quantified as the time from the end of inflation until the equation of state makes transition from the value $w_{re}$ it had during reheating to $w=1/3$, which is the beginning of radiation domination. 
\begin{figure}[tbp]
    \centering
    \includegraphics[width=13cm]{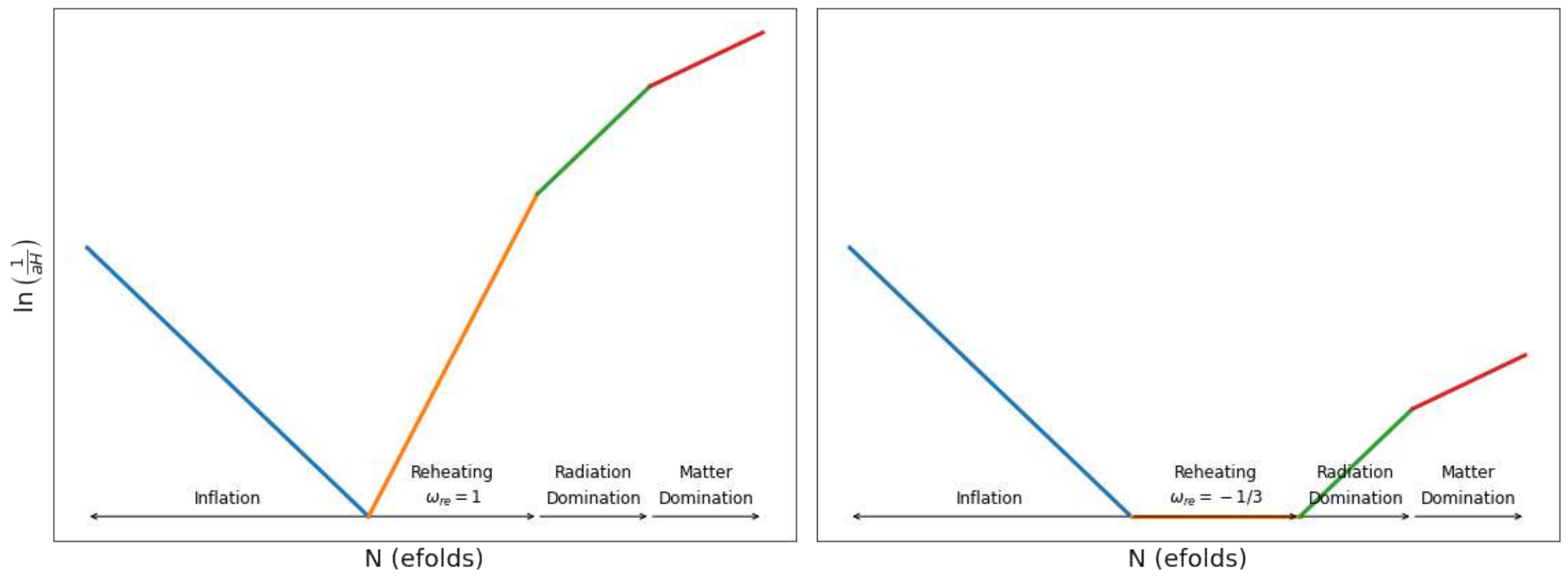}
    \caption{Evolution of the co-moving horizon distance over time. The left plot is with assumption that $\omega_{re}=1$ and right one represents $\omega_{re}=-1/3$.}
    \label{fig:horizon}
\end{figure}
\noindent In figure \ref{fig:horizon} we have shown how the co-moving horizon evolves during different phases of time. One can see that the modes that enter back into the horizon are dependent on how long the reheating lasts i.e $N_{re}$ and the equation of state parameter $\omega_{re}$ during reheating. the smaller $\omega_{re}$ is during reheating, later the modes re-enter the horizon, and more efolds will be necessary in the post-inflation period.\\
 In general, the duration
\begin{equation}
    N_{re}=ln\left(\frac{a_{re}}{a_{end}}\right)=\frac{1}{3(1+\omega_{re})}ln \left(\frac{\rho_{end}}{\rho_{re}} \right)
    \label{Nre}
\end{equation}  
and the energy density here can be written as
\begin{equation}
    \rho_{re}=\frac{\pi^2g_{re}}{30}T_{re}^4
    \label{rho}
\end{equation}
with $g_{re}$ being the relativistic degrees of freedom. With entropy conservation between today and at the time of reheating we obtain \cite{Cook}
\begin{equation}
    T_{re}=\left ( \frac{43}{11g_{re}}  \right )^{1/3} \left(\frac{a_0T_0}{k}\right)H_ke^{-N_k}e^{-N_{re}}
    \label{Tre}
\end{equation}

Using Eq.\ref{Nre}, Eq.\ref{rho} and Eq.\ref{Tre} and the fact that $\rho_{end}=\frac{3}{2}V_{end}$ when we substitute $w=1/3$, obtain the number of e-folds of reheating as
\begin{equation}
    N_{re}=\frac{4}{1-3\omega_{re}}\left[61.6-ln\left(\frac{V_{end}^{1/4}}{H_k}\right)-N_k\right]
    \label{Nre2}
\end{equation}

\section{Slow-Roll Analysis} \label{Slow-Roll Analysis}
Here we work out a model of inflation with a potential of the form
\begin{equation}
    V = V_0\cdot \left [ tan^{-1}\left(\frac{\kappa \phi}{m_p}\right) \right ]^2
    \label{pot}
\end{equation}
where $\kappa$ is a dimensionless free parameter and $V_0$ has dimensions of $[Energy]^4$.  The potential slow-roll parameters are found to be 
\begin{equation}
    \epsilon=\frac{4\kappa^2}{2(1+\kappa^2 x^2)^2[tan^{-1}(\kappa x)]^2} \label{ep}, \quad \eta=\frac{2\kappa^2}{(1+\kappa^2 x^2)^2tan^{-1}(\kappa x)}\left[\frac{1}{tan^{-1}(\kappa x)}-2\phi \right] \quad \text{where} \quad x=\frac{\phi}{m_p}
\end{equation}
\noindent and the amplitude of the scalar power spectrum is calculated as

\begin{equation}
    A_s=\frac{1}{48\pi^2}\frac{V_0tan^{-1}(kx)^4(m_p^2+k^2\phi^2)}{k^2m_p^8}
\end{equation}
Since inflation ends when $\epsilon=1$ we can find the value of $\phi$ at the end of inflation by substituting $\epsilon=1$ in Eq. \ref{ep} and solve for $\phi$. Since its a transcendental equation, we solve it numerically to find $\phi_e$.
We compute the integral in Eq. \ref{Ne} as

\begin{equation}
    N_k=\frac{1}{12\kappa^2} \left[ -\kappa^2x^2-2log(\kappa^2x^2+1)+2(\kappa^2x^2+3)\kappa x \cdot tan^{-1}(\kappa x) \right]^{x_{max}}_{x_{min}}
\end{equation}
where $x_{max}=\frac{\phi_k}{m_p}$ and $x_{min}=\frac{\phi_e}{m_p}$.

This equation can be inverted to find $\phi$ at N e-folds before the end of inflation when a mode \textbf{'k'} exited the horizon. Again this is done numerically for $N=40$ to $N=60$. Using this we find numerical values for our slow-roll parameters at a particular value of $N$, which are then used to calculate spectral index $n_s$,tensor-to-scalar ratio $r$ and running of spectral index $\frac{d n_s}{d \ln k}$ using Eq. \ref{si}.

\begin{figure}[tbp]
    \centering
    \includegraphics[width=10.5cm]{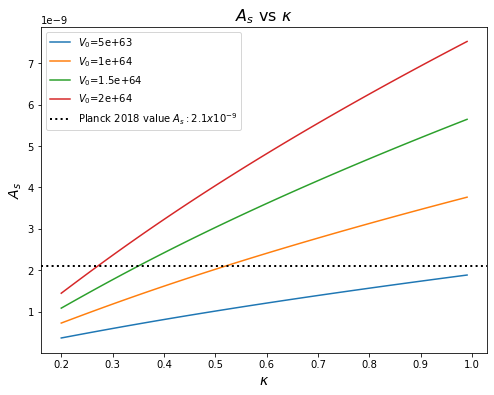}
    \caption{Plot of Amplitude of scalar power spectrum $A_s$ vs $\kappa$}
    \label{Plot}
\end{figure}

In table \ref{tab:para}, we present values of these quantities for some values of free parameter $\kappa$.
\begin{table}[htb]
    \centering
     \addtolength{\tabcolsep}{-0.5pt}
        \small
    \begin{tabular}{|c c c c c c c|}
    \hline
    $\kappa=0.05$ &&&&&&\\
    \hline
     $\phi_e/m_p$ & $\phi_*/m_p$ & $N$ & $\epsilon$ & $\eta$ & $n_s$ &$r$ \\
     1.410 & 12.049 & 40 & 0.009 & 0.003 & 0.951 & 0.146\\
     1.410 & 13.311 & 50 & 0.007 & 0.002 & 0.961 & 0.111 \\
     1.410 & 14.425 & 60 & 0.006 & 0.001 & 0.968 & 0.089\\
    \hline\hline
    $\kappa=0.2$ &&&&&&\\
    \hline
    $\phi_e/m_p$ & $\phi_*/m_p$ & $N$ & $\epsilon$ & $\eta$ & $n_s$ &$r$ \\
    1.350 & 9.297 & 40 & 0.003 & -0.01 & 0.958 & 0.056\\
    1.350 & 10.066 & 50 & 0.003 & -0.009 & 0.967 & 0.041\\
    1.350 & 10.735 & 60 & 0.02 & -0.008 & 0.773 & 0.032\\
    \hline\hline
        $\kappa=0.5$ &&&&&&\\
    \hline
    $\phi_e/m_p$ & $\phi_*/m_p$ & $N$ & $\epsilon$ & $\eta$ & $n_s$ &$r$ \\
    1.166 & 7.011 & 40 & 0.002 & -0.014 & 0.963 & 0.027\\
    1.166 & 7.550 & 50 & 0.001 & -0.011 & 0.970 & 0.020\\
    1.166 & 8.019 & 60 & 0.001 & -0.009 & 0.975 & 0.016\\
    \hline
    
    \end{tabular}
    \caption{Calculated values of $n_s,r$ for different values of free parameter $\kappa$ across different $N$}
    \label{tab:para}
\end{table}
%\newpage
To understand the nature of Primordial spectrum generated by our model, we plot the graph of Amplitude of scalar power spectrum vs $\kappa$ in figure \ref{Plot}. The value of the amplitude of the scalar power is given as $A_s \simeq {2.1} \times10^{-9}$ by the \emph{Planck} 2018 results. For our analysis we take the amplitude of the potential $V_0$ to be of the order of $10^{64}~GeV^4$ and set the number of e-folds at $N=55$. We find that for all potential amplitudes except $V_0=0.5\times10^{64}$ the values of $\kappa$ for which the power spectrum amplitude 
$A_s$ agrees with the \emph{Planck} 2018 measurement fall within the range 
$\kappa\in [0.270,0.520]$. In what follows, we will see that other observable quantities also agree with the \emph{Planck} 2018 and ACT collaboration results within this same $\kappa$-range.

\begin{figure}[tbp]
    \centering
\begin{subfigure}{0.48\textwidth}
    \includegraphics[width=\linewidth]{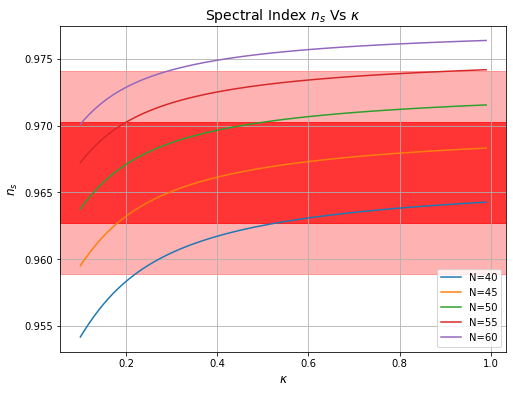}
    
  \end{subfigure}
  \hfill
 \begin{subfigure}{0.48\textwidth}
    \includegraphics[width=\linewidth]{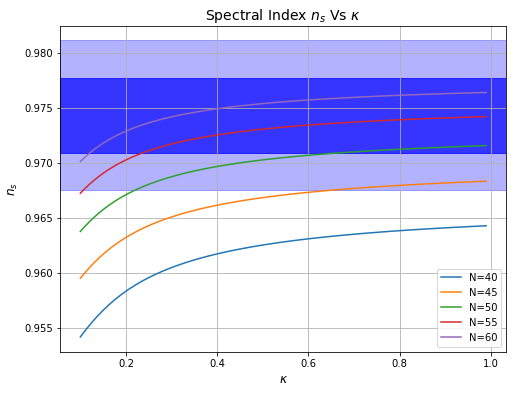}
    \end{subfigure}
  \caption{ \label{Plot1}{Graph of spectral index $n_s$ vs $\kappa$ for various $N$. The dark and the light regions correspond to the 1$\sigma$ and 2$\sigma$ bounds on $n_s$ from \emph{Planck} 2018 (left) and ACT (right) data respectively.}}
\end{figure}

In figure \ref{Plot1}, to illustrate the behavior of spectral index $n_s$ against change in $\kappa$ we show a graph of $n_s$ vs $\kappa$ for different e-folds of inflation $N$. We have also included the constraints on $n_s$ from \emph{Planck} (left) as well as ACT collaboration (right). For smaller values of $\kappa$, $n_s$ increases and then starts plateauing for higher values. As illustrated in the figure for N=60, there is a notable tension between the two datasets. The analysis by the \emph{Planck} 2018 collaboration imposes a stringent upper limit, disfavoring $\kappa>0.4$. In contrast, the data from the ACT collaboration accommodate, and may even prefer, a higher range of $\kappa>0.5$ to satisfy the spectral index measurement. There is a significant tension in the model's strong dependence on the number of e-folds, $N$. The \emph{Planck} 2018 data favor intermediate values around $N\simeq 50-55$, largely excluding the $N=60$ case for most values of $\kappa$. In direct contrast, the ACT data, which measure a systematically higher spectral index ($n_s$), decisively prefer $N\geq 60$ and rule out the lower N values favored by \emph{Planck}.

\begin{figure}[tbp]
 \centering
  \includegraphics[width=12.5cm]{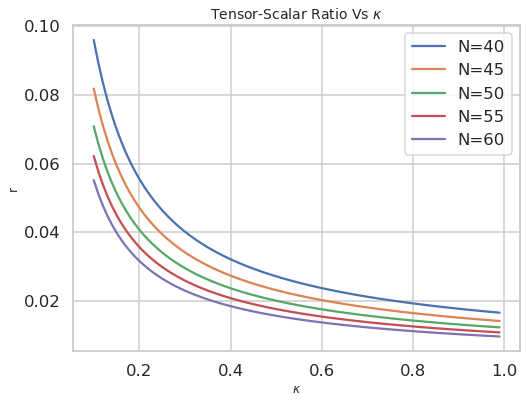}
  \caption{ \label{Plot2}{Graph of tensor-to-scalar ratio $r$ vs $\kappa$ for various $N$}}
\end{figure}

Next in figure \ref{Plot2} we see the variation of the tensor-to-scalar ratio $r$ with our parameter $\kappa$. \emph{Planck} 2018 TT,TE,EE+lowE+lensing+BAO data put a 2$\sigma$ bound of $r<0.066$ on the tensor-to-scalar ratio, whereas P-ACT-LB-BK18 puts $2\sigma$ bound of $r<0.038$. When compared to our model we see that our model obeys this upper bound for $\kappa>0.16$. For $N=50$, $r$ is found to obey the upper bound for $\kappa>0.11$. These two plots on $n_s$ and $r$ will be useful in constraining the free parameter $\kappa$ when combined with the \emph{Planck} 2018 and ACT collaboration data. 

\begin{figure}[tbp]
 \centering
  \includegraphics[width=11.5cm]{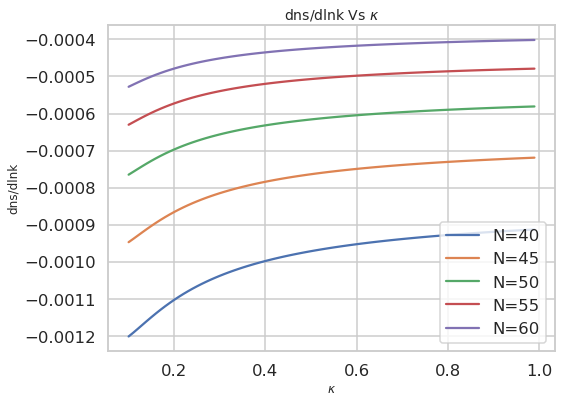}
  \caption{ \label{Plot3}{Graph of running of spectral index $\frac{dn_s}{dlnk}$ vs $\kappa$ for various $N$}}
\end{figure}

\begin{figure}[tbp]
    \centering
    \includegraphics[width=15cm]{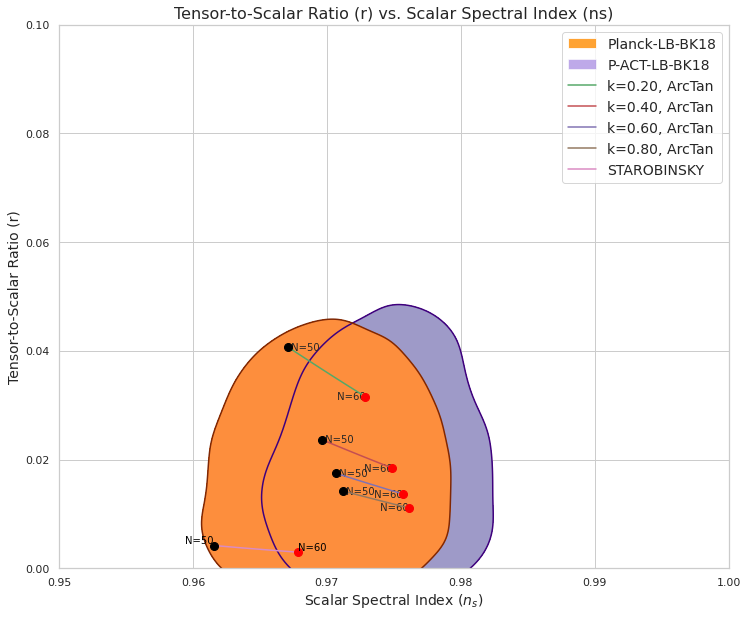}    \caption{Tensor-to-scalar ratio $r$ versus scalar spectral index $n_s$ for the inverse-tangent potential with $\kappa = 0.20, 0.40, 0.60,$ and $0.80$, shown for $N=50$ and $N=60$ e-folds. The shaded contours represent the $2\sigma$ bounds from \emph{\emph{Planck}}-LB-BK18 (orange) and P-ACT-LB-BK18 (purple) datasets.}
    \label{fig:nsr}
\end{figure}

To understand how spectral index changes with changing scales we define a parameter called running of spectral index $\alpha=\frac{dn_s}{d lnk}$. In figure \ref{Plot3} of running of spectral index vs $\kappa$ we see $\alpha$ is always less than $-0.0004$ and has a minimum value of -0.0012 for $\kappa=0.1$ and $N=40$. Again if we compare with the \emph{Planck} 2018 and ACT collaboration results, they put it a 1$\sigma$ bound on $\alpha$ as $\alpha=-0.0041 \pm 0.0067$ and $\alpha=0.0062\pm0.0052$. Our results are in excellent agreement with the \emph{Planck} collaboration results. Constraints from \emph{Planck} prefer a slight negative running of the spectral index but ACT constraints disfavor negative running of spectral index.
To better understand our results and to make a comparison with recent CMB constraints, we show the $r$ vs. $n_s$ plot in figure~\ref{fig:nsr}. In this figure, we present our predictions for the inverse-tangent potential for different values of the parameter $\kappa$, specifically $\kappa = 0.20, 0.40, 0.60,$ and $0.80$, with results displayed for $N=50$ and $N=60$ e-folds. For comparison, we also plot the predictions of the Starobinsky potential, one of the most successful inflationary models. The shaded contours represent the current observational bounds: the orange region corresponds to the \emph{Planck}-LB-BK18 dataset, while the purple region represents the P-ACT-LB-BK18 dataset. From figure~\ref{fig:nsr}, it is evident that the predictions from the inverse-tangent potential lie well within the $2\sigma$ bounds of both \emph{Planck} and ACT data. Increasing the free parameter $\kappa$ leads to a suppression of the tensor-to-scalar ratio $r$ while slightly shifting the scalar spectral index $n_s$ to higher values. Interestingly, while the Starobinsky model is in excellent agreement with the\emph{Planck} data, it lies outside the ACT bounds. In contrast, the inverse-tangent potential remains consistent with both datasets, suggesting that the inverse-tangent potential could serve as a promising alternative to Starobinsky's inflation in light of the latest CMB observations.

\begin{table}[htb]
    \centering
     \addtolength{\tabcolsep}{-0.5pt}
        \small
    \begin{tabular}{|c|c| c| c| c |}
    \hline
    Potential & $n_s(N=50)$& $n_s(N=60)$& $r(N=50)$&$r(N=60)$\\

    \hline
    Starobinsky & 0.9616&0.9678&0.0042&0.0030\\
    Inverse-tangent ($\kappa=0.2$) & 0.9671&0.9729&0.0407&0.0316\\
    Inverse-tangent ($\kappa=0.4$) & 0.9697&0.9749&0.0236&0.0184\\
    Inverse-tangent ($\kappa=0.6$) & 0.9707&0.9757&0.0175&0.0137\\
    Inverse-tangent ($\kappa=0.8$) & 0.9712&0.9761&0.0142&0.0111\\
    \hline
    
\end{tabular}
    \caption{Calculated values of $n_s,r$ for different values of free parameter $\kappa$ across different $N$ compared with Starobinsky's potential}
    \label{tab:RAT}
\end{table}

\begin{figure}[tbp]
    \centering
    \includegraphics[width=11.5cm]{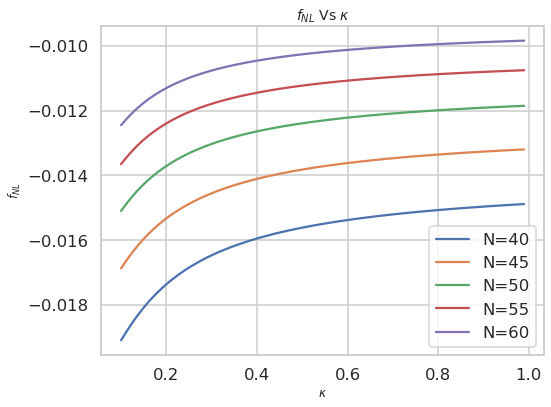}
    \caption{Plot of $f_{NL}$ vs $\kappa$}
    \label{fig:Plot4}
\end{figure}

 We next study the non-Gaussianity of the CMB power spectrum. From figure \ref{fig:Plot4} we confirm the suppression of non-Gaussianity in the power spectrum \cite{Planck2018NG}. The value of $f_{NL}$ is of the order of $f_{NL}\sim O(\epsilon,\eta)$. We see that $f_{NL}$ increases initially for small values of $\kappa$ but more or less remains constant for higher values. Also the non-Gaussianity is seen to be higher for more number of e-folds $N$. It varies in the range $-0.0191$ to $-0.0098$ for $N=40$ to $N=60$.

 %%%%%%%%%%%%%%%%%%%%%%%%%%%%%%%%%%%%%%%%%%%
\section{Reheating Analysis}\label{Reheating Analysis}
In the reheating analysis, we want to calculate the duration of reheating $N_{re}$ and the thermalization temperature $T_{re}$. As stated earlier in our analysis we have assumed constant equation of state parameter $\omega_{re}$ throughout the reheating process. First we need to find $H_k$, we do so by using the definition of tensor-to-scalar ratio $r=\frac{P_t}{P_R}$ and $r=16\epsilon$. Hence, we get
\begin{equation}
    H_k=\pi m_p \sqrt{8A_s\epsilon_k}.
\end{equation}
We use $\phi_e$ from the slow-roll analysis to calculate $V_{end}$. Also we need to obtain $\phi(n_s)$ which can be done by inverting Eq. \ref{si} for $n_s$ in terms of $\phi$, once we  find $\phi(n_s)$ we then use it to calculate $N_{re}$ and $T_{re}$ as a function of the spectral index $n_s$.

\begin{figure}[tbp]
    \centering
\begin{subfigure}{0.48\textwidth}
    \includegraphics[width=\linewidth]{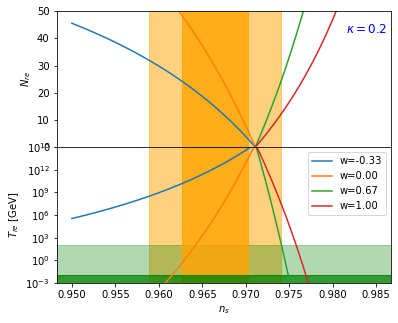}
    
  \end{subfigure}
  \hfill
 \begin{subfigure}{0.48\textwidth}
    \includegraphics[width=\linewidth]{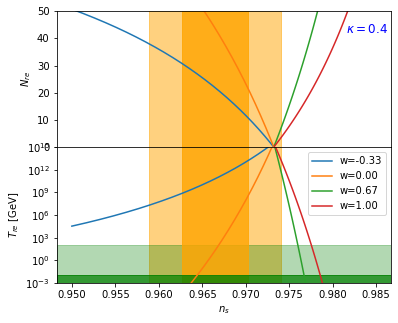}
    
  \end{subfigure}
  \hfill

  \begin{subfigure}{0.48\textwidth}
    \includegraphics[width=\linewidth]{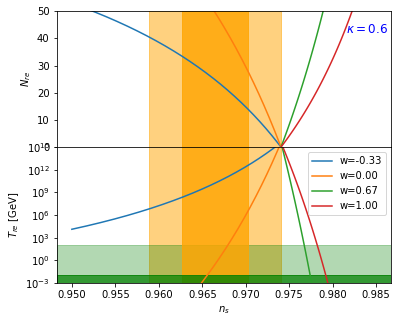}
    
  \end{subfigure}
  \hfill
  \begin{subfigure}{0.48\textwidth}
    \includegraphics[width=\linewidth]{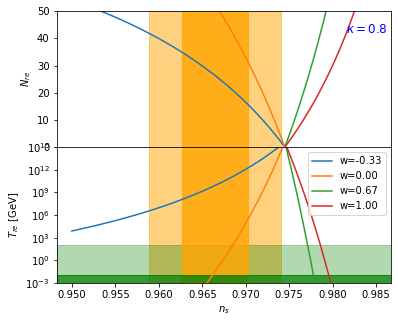}
    
  \end{subfigure}
  \hfill
    \caption{Plot of $T_{re}$ and $N_{re}$ for different values of free parameter $\kappa$.
    Dark and light shaded regions correspond to $1\sigma$ and $2\sigma$ bounds on $n_s$ from \emph{Planck} 2018 collaboration.}
    \label{fig:Tre_Nre_A}
\end{figure}

\begin{figure}[tbp]
    \centering
\begin{subfigure}{0.48\textwidth}
    \includegraphics[width=\linewidth]{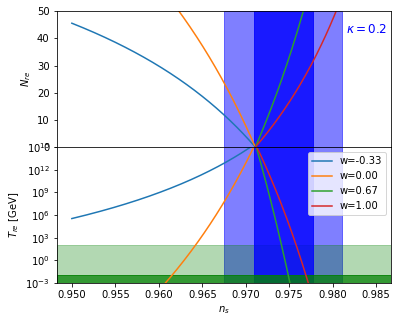}
    
  \end{subfigure}
  \hfill
 \begin{subfigure}{0.48\textwidth}
    \includegraphics[width=\linewidth]{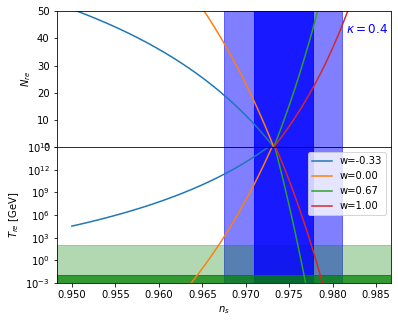}
    
  \end{subfigure}
  \hfill

  \begin{subfigure}{0.48\textwidth}
    \includegraphics[width=\linewidth]{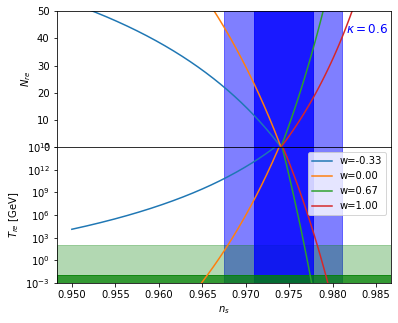}
    
  \end{subfigure}
  \hfill
  \begin{subfigure}{0.48\textwidth}
    \includegraphics[width=\linewidth]{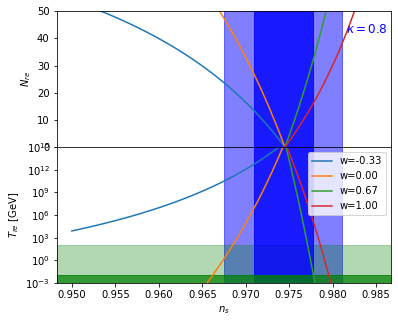}
    
  \end{subfigure}
  \hfill
    \caption{Plot of $T_{re}$ and $N_{re}$ for different values of free parameter $\kappa$. Dark and light shaded regions correspond to $1\sigma$ and $2\sigma$ bounds on $n_s$ from ACT collaboration.}
    \label{fig:Tre_Nre}
\end{figure}

In figure \ref{fig:Tre_Nre_A} \& \ref{fig:Tre_Nre}  we have plotted the predictions of $T_{re}$ and $N_{re}$ for different values of $\kappa$. 
Each panel plots reheating predictions for the inverse-tangent inflationary potential as a function of the scalar spectral index $n_s$. The two plotted quantities are the number of e-folds of reheating $N_{re}$ (shown on the upper part / linear axis) and the reheating temperature $T_{re}$ (shown on the lower part / logarithmic axis). Curves of different color correspond to different assumed constant reheating equations of state $\omega_{re}=\{-1/3,0,2/3,1\}$. The shaded horizontal/vertical bands are observational or physical bounds: orange = \emph{Planck} $n_s$ region, purple = ACT $n_s$ region; dark/ light green bands mark the BBN lower limit \cite{Kawasaki,Kawasaki2} and the electroweak scale.  
Since \emph{\emph{Planck}}'s preferred $n_s$ region sits at slightly lower value than ACT, for the shown $\kappa$ values higher $\omega_{re}$ curves (2/3 and 1) are outside the $2\sigma$ band thus \emph{Planck} disfavors high $w_{re}$ reheating histories. It is evident from the graph that a higher temperature corresponds to lesser number of e-folds. From the plots we see that as the value of the free parameter increases the graph shifts towards higher values of $n_s$. We try to constrain the reheating temperature $T_{re}$ and Number of e-folds of reheating with the help of $1\sigma$ bound on $n_s$. When we refer to \emph{Planck} data, for $\kappa=0.2$ the values of $T_{re}$ that lie within the 1$\sigma$ bound of $n_s$ lie in the range $10^{10}$ GeV to $10^{14}$ GeV while $N_{re}$ is in the range 24 to 3 with the higher value corresponding to a lower temperature for $\omega_{re}=-1/3$. For $\omega_{re}=0$ this temperature range gets broader with $T_{re}$ ranging from $10^{-1}$ GeV to $10^{13}$ GeV. And $N_{re}$ varying from 48 to 6, again with the higher value corresponding to a lower temperature.

And as we increase $\kappa$ we see that the upper bound on $T_{re}$ decreases as well as the lower bound for both the eos parameter cases. For $\kappa=0.4$ and eos parameter $\omega_{re}=0$, the predicted values of $T_{re}$ lies below $10^8$ GeV which is interesting since this is well below the temperatures which may introduce back the monopole problem and production of other massive relics such as gravitinos. Although the lower bound obtained is below the BBN constraint but for $\kappa=0.4$, if we consider $n_s>0.9638$ then $T_{re}>1$ MeV. We use this to constrain $N_k$ which determines the solution to horizon and flatness problem with the help of lower bound on $T_{re}$. With our analysis we observe $N_k>42$ which is in agreement with our slow roll estimation of $N_k$ ranging from 50-60. 
A slightly different behavior is seen when we consider ACT constraints on $n_s$. Since ACT prefers slightly higher $n_s$ the curves $\omega=2/3$ and $\omega=1$ which were excluded by \emph{Planck} are now well within the allowed region for several $\kappa$ values. Which makes kination-like histories more compatible with the inverse-tangent model. Since high $\omega_{re}$ can accommodate lower temperatures ($\approx10^6GeV$) this analysis is safe with respect to monopole/gravitino relic problems. It is clear that while ACT allows high $\omega_{re}$ solutions but still the parameter space intersects with the green region; and those particular points need to be excluded. For $\kappa=0.2$ only reheating temperatures of the order of $10^{15}$ GeV are allowed by ACT for $\omega_{re}=-1/3$ and $\omega_{re}=0.$. Whereas for the case of $\omega_{re}=2/3$ the range becomes much wider as the temperature can range from $10^-{19}$ GeV to $10^{15}$ GeV. A similar pattern pattern is observed for the case of $\omega_{re}=1$

We present our results in Table \ref{tab:Tre_P} and \ref{tab:Tre_A} in a more compact form.

\begin{table}[tbp]
\centering
\caption{Order of magnitude bounds on $T_{re}$ and $N_{re}$}
\renewcommand{\arraystretch}{1.2}

\begin{subtable}[b]{0.45\textwidth}
\centering
\begin{adjustbox}{max width=\textwidth}
\begin{tabular}{|c c c c c|}
\hline
\multicolumn{5}{|c|}{\boldmath$\kappa = 0.4$} \\
\hline
$\omega_{re}$ & $T_{re}$ (min) & $T_{re}$ (max) & $N_{re}$ (max) & $N_{re}$ (min) \\
\hline
$-1/3$ & $10^{8}$   & $10^{13}$ & 30 & 11 \\
0      & $10^{-3}$  & $10^{8}$  & 61 & 21 \\
\hline\hline

\multicolumn{5}{|c|}{\boldmath$\kappa = 0.8$} \\
\hline
$\omega_{re}$ & $T_{re}$ (min) & $T_{re}$ (max) & $N_{re}$ (max) & $N_{re}$ (min) \\
\hline
$-1/3$ & $10^{8}$   & $10^{12}$ & 35 & 16 \\
0      & $10^{-7}$  & $10^{5}$  & 69 & 31 \\
\hline
\end{tabular}
\end{adjustbox}
\caption{Bounds based on $1\sigma$ Planck 2018 constraint on $n_s$.}
\label{tab:Tre_P}
\end{subtable}
\hfill
\begin{subtable}[b]{0.48\textwidth}
\centering
\begin{adjustbox}{max width=\textwidth}
\begin{tabular}{|c c c c c|}
\hline
\multicolumn{5}{|c|}{\boldmath$\kappa = 0.4$} \\
\hline
$\omega_{re}$ & $T_{re}$ (min) & $T_{re}$ (max) & $N_{re}$ (max) & $N_{re}$ (min) \\
\hline
$-1/3$ & $10^{13}$ & $10^{15}$ & 8  & 0 \\
0      & $10^{10}$ & $10^{15}$ & 17 & 0 \\
$2/3$  & $10^{-8}$ & $10^{15}$ & 43 & 0 \\
1      & $10^{1}$  & $10^{15}$ & 22 & 0 \\
\hline\hline

\multicolumn{5}{|c|}{\boldmath$\kappa = 0.8$} \\
\hline
$\omega_{re}$ & $T_{re}$ (min) & $T_{re}$ (max) & $N_{re}$ (max) & $N_{re}$ (min) \\
\hline
$-1/3$ & $10^{12}$ & $10^{15}$ & 13 & 0 \\
0      & $10^{7}$  & $10^{15}$ & 26 & 0 \\
$2/3$  & $10^{-2}$ & $10^{15}$ & 32 & 0 \\
1      & $10^{5}$  & $10^{15}$ & 16 & 0 \\
\hline
\end{tabular}
\end{adjustbox}
\caption{Bounds based on $1\sigma$ ACT constraint on $n_s$.}
\label{tab:Tre_A}
\end{subtable}

\end{table}

\begin{figure}[tbp]
    \centering
\begin{subfigure}{0.48\linewidth}
    \includegraphics[width=\linewidth]{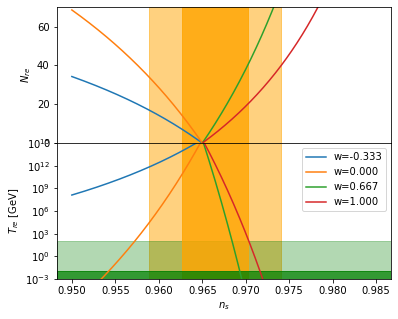}
    
  \end{subfigure}
  \hfill
 \begin{subfigure}{0.48\linewidth}
    \includegraphics[width=\linewidth]{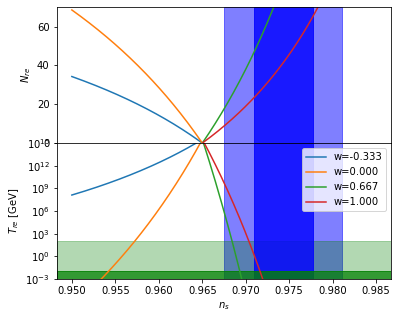}
    
  \end{subfigure}
  \caption{Plot of $T_{re}$ and $N_{re}$ for Starobinsky's potential. The figure on left corresponds to $1\sigma$ and $2\sigma$ bounds on $n_s$ from \emph{Planck} 2018 Collaboration whereas figure on right corresponds to constraints on $n_s$ from ACT  Collaboration. }
\label{R_tr}
\end{figure}

We now compare our results from Reheating with the well known Starobinsky's potential. In figure \ref{R_tr} we plot a graph similar to figure \ref{fig:Tre_Nre_A} \& \ref{fig:Tre_Nre}. If we take \emph{Planck}'s data for reference (left side), for the case $\omega_{re}=-1/3$ the lower bound on $T_{re}$ is $10^{14}$ GeV and for $\omega_{re}=0$ it is $10^{11}$ GeV. When $\omega_{re}=2/3$ we observe that for $n_s<0.9695$ the value of $T_{re}$ lies above the BBN constraint of 1 MeV. In the case of $\omega_{re}=1$ the lower bound on $T_{re}$ is $10^{2}$ GeV. 
When the recent ACT constraints on $n_s$ is taken into account $\omega_{re}=-1/3$ and $\omega_{re}=0$ lie well outside the $2\sigma$ allowed region. Although for lower reheating temperatures ($>10^9GeV$) $\omega_{re}=2/3$ and $\omega_{re}=1$ cases fall well within the bound. We see that for all the eos parameter cases the lower bound on $N_{re}$ is zero, which corresponds to instantaneous reheating. From this analysis, $w_{re}=2/3$ and $w_{re}=1$ seem the most physically viable cases. 

Finally, in Table \ref{tab:Tre_R} and \ref{tab:Tre_RP}(presented below), we have presented the maximum and minimum bounds on $T_{re}$ and $N_{re}$, we obtained with Starobinsky's potential.
\begin{table}[tbp]
\centering
\caption{Order of magnitude bounds on $T_{re}$ and $N_{re}$ for Starobinsky inflation.}
\vspace{0.3cm}

\begin{subtable}[t]{0.48\linewidth}
\centering
\begin{adjustbox}{max width=\textwidth}
\begin{tabular}{|c|c|c|c|c|}
\hline
$\omega_{re}$ & $T_{re}$ (min) & $T_{re}$ (max) & $N_{re}$ (max) & $N_{re}$ (min) \\
\hline
$-1/3$ & $10^{14}$ & $10^{15}$ & 7  & 0 \\
0      & $10^{11}$ & $10^{15}$ & 14 & 0 \\
$2/3$  & $10^{-7}$ & $10^{15}$ & 41 & 0 \\
1      & $10^{2}$  & $10^{15}$ & 20 & 0 \\
\hline
\end{tabular}
\end{adjustbox}
\caption{Bounds based on $1\sigma$ Planck 2018 constraint on $n_s$.}
\label{tab:Tre_R}
\end{subtable}
\quad
\begin{subtable}[t]{0.48\linewidth}
\centering
\begin{adjustbox}{max width=\textwidth}
\begin{tabular}{|c|c|c|c|c|}
\hline
$\omega_{re}$ & $T_{re}$ (min) & $T_{re}$ (max) & $N_{re}$ (max) & $N_{re}$ (min) \\
\hline
$2/3$ & -- & $10^5$ & -- & 18 \\
1     & -- & $10^9$ & -- & 9  \\
\hline
\end{tabular}
\end{adjustbox}
\caption{Bounds based on $2\sigma$ ACT constraint on $n_s$.}
\label{tab:Tre_RP}
\end{subtable}

\end{table}

%%%%%%%%%%%%%%
%In Table \ref{tab:Tre_R} and \ref{tab:Tre_RP} we have presented the maximum and minimum%bounds on $T_{re}$ and $N_{re}$, we obtained with Starobinsky's potential. 
When ACT collaboration results were used we found that temperatures are too low if we use $1\sigma$ bound on $n_s$ hence we have used $2\sigma$ bound. Also the lower bound even in that case is very low and hence we only show an upper bound on $T_{re}$ and $N_{re}$. \\

\noindent {\bf Comparing our model with the model of Starbinsky}: We observe that while the Starobinsky model consistently predicts high reheating temperatures and short reheating durations, {\it our model allows significantly longer reheating phases and lower reheating temperatures, including scenarios where $T_{re}\sim O(MeV)$, which is close to the BBN limit when we use Planck's results}.
These results imply that our model is more flexible in accommodating non-standard thermal histories, such as those relevant for late-time baryogenesis, suppression of relic production (e.g., gravitinos or moduli), or delayed thermalization.

\section{Conclusion}\label{Conclusion}
We employed our inverse-tangent potential $ V = V_0\cdot \left [ tan^{-1}(\frac{\kappa \phi}{m_p})\right ]^2$ to the domain of slow-roll inflationary expansion. We have used \emph{Planck} 2018 results on cosmological parameters and ACT collaboration results to constrain the value of free parameter $\kappa$ in our model. We found that for $\kappa$ ranging from 0.2 to 0.8 the value of the spectral index $n_s$ is well within the 2$\sigma$ bound of both the \emph{Planck} as well as the ACT collaboration data. The value of tensor-to-scalar ratio is also found to be within the allowed upper bound for the same $\kappa$ range. We used the amplitude of scalar power spectrum to determine the energy-scale of inflation, which we found to be of the order of $10^{16}$ GeV. Running of the spectral index was found to be in the range -0.0004 to -0.0012. The non-Gaussianity parameter calculated for this model was found to be negligible and of the order of $f_{NL}\sim O(\epsilon,\eta)$, which is what is expected in the case of single-field inflation. Therefore from our analysis we infer that our inverse-tangent model is fully compatible with the predictions of the \emph{Planck} 2018 and ACT results for $\kappa \in [0.2,0.8]$. Hence our inverse-tangent potential can serve as a viable alternative to standard inflationary potentials like Starobinsky which are now excluded by ACT at $2\sigma$ level. \\
~By using entropy conservation we have determined the reheating temperature $T_{re}$ and the duration of reheating $N_{re}$. We find that the lower bound on $T_{re}$ can be of the order of MeV and thus requires a very high number of e-foldings of reheating. We then extended our reheating analysis to Starobinsky model and compared our results to the Starobinsky model. Our potential predicts a broader range of reheating scenarios, including viable low-temperature reheating phases. This makes it compatible with a wider range of post-inflationary physics.
%\newpage

\section{Acknowledgment}
The MA thanks the BITS Pilani K K Birla Goa campus for the fellowship support. The work of PKD is supported by ANRF Grant No. CRG/2023/008877.

\end{document}